\documentclass[final,3p,times,12pt]{elsarticle}

\usepackage{graphicx,graphics}
\usepackage{amssymb,epsfig,amsmath,comment,hyperref,verbatim}
\usepackage{color}
\begin{document}

\begin{frontmatter}
\journal{Journal of Voice}

\title{Can a Machine Distinguish High and Low Amount of Social Creak in Speech?}

\author{Anne-Maria Laukkanen$^1$, Sudarsana Reddy Kadiri$^2$, Shrikanth Narayanan$^2$, and Paavo Alku$^3$}
\address{$^1$Speech and Voice Research Laboratory, Tampere University, Finland\\$^2$Signal Analysis and Interpretation Laboratory (SAIL), University of Southern California, USA\\$^3$Department of Information and Communications Engineering, Aalto University, Finland}

\cortext[mycorrespondingauthors]{Corresponding author (Sudarsana Reddy Kadiri: skadiri@usc.edu and Paavo Alku: paavo.alku@aalto.fi)}

\begin{abstract}
\textbf{Objectives}: 
Increased prevalence of social creak particularly among female speakers has been reported in several studies. The study of social creak has been previously conducted by combining perceptual evaluation of speech with conventional acoustical parameters such as the harmonic-to-noise ratio and cepstral peak prominence. In the current study, machine learning (ML) was used to automatically distinguish speech of low amount of social creak from speech of high amount of social creak.\\
\textbf{Methods}:
The amount of creak in continuous speech samples produced in Finnish by 90 female speakers was first perceptually assessed by two voice specialists. Based on their assessments, the speech samples were divided into two categories (low $vs$. high amount of creak). Using the speech signals and their creak labels, seven different ML models were trained. Three spectral representations were used as feature for each model. \\ 
\textbf{Results}:
The results show that the best performance (accuracy of 71.1\%) was obtained by the following two systems: an Adaboost classifier using the mel-spectrogram feature and a decision tree classifier using the mel-frequency cepstral coefficient feature. \\ 
\textbf{Conclusions}: The study of social creak is becoming increasingly popular in sociolinguistic and vocological research. The conventional human perceptual assessment of the amount of creak is laborious and therefore ML technology could be used to assist researchers studying social creak.  The classification systems reported in this study could be considered as baselines in future ML-based studies on social creak. 

\end{abstract}

\end{frontmatter}

\section{Introduction}
Creaky voice refers to a raspy, rough sounding voice that is often low-pitched. 
Keating et al. (\cite{keating2015acoustic}) classified creaky voice into seven sub-types: prototypical creaky voice, vocal fry, multiply- pulsed voice, aperiodic voice, tense or pressed voice and non-constricted creak. Although creaky voice shares perceptual and acoustic similarities with dysphonia (such as perceptual roughness, strain or markedly low pitch, as well as irregular, multiple pulsed and even chaotic signal structure, see e.g., \cite{hirano1981clincal}, \cite{dejonckere2001basic}), it is a variant of voice use, and all the above mentioned subtypes of creaky voice appear in healthy speakers (\cite{keating2015acoustic}, \cite{laukkanen2021relations}. Creaky voice appears very often in sentence endings, and it can be used to signal phrase ending or turn taking in a conversation (\cite{kreiman1982perception}, \cite{henton1988creak},  \cite{redi2001variation}, \cite{ogden2001turn}). In some languages, it is used to create phonemic contrasts (\cite{abercrombie1971elements}, \cite{laver1994principles}, \cite{gordon2001phonation}). Furthermore, it may be used to express emotions and attitudes, e.g., it has been related to expression of hesitance, complaining, boredom, relaxation, intimacy, contentedness (\cite{cantor2017vocal}, \cite{gobl2003role}) or anger of low-activation (\cite{cullen2013creaky}). 
In addition to specific linguistic and emotional expression use, creaky voice appears to be a general sociolinguistic marker. It may appear in all sentence positions, and it is used in different age groups (\cite{oliveira2016comparison}) and as well in professional (radio) speakers as in untrained speakers (\cite{redi2001variation}). More or less continuous creaky voice may be used as a marker of social status and authority (\cite{yuasa2010creaky})  or belonging to a certain social group (\cite{mendoza2011semiotic}). It may also be a habit, seemingly without any particular role as a marker, unless the role is then to signal informality, like it seems to be the case in Estonian speakers (\cite{aare2014creaky}). Creaky voice may be a preferred choice also because it is possible to establish using a low subglottic pressure, at least the subtype ‘vocal fry’ (\cite{blomgren1998acoustic}). 
In this study, we name all creak in healthy speakers' voices as ‘$social$ $creak$’. Due to the similarities between creaky voice and characteristics of dysphonia, continuous creaky voice use may be even called as social/voluntary dysphonia.

The prevalence of social creak 
has increased remarkably (\cite{yuasa2010creaky}, \cite{hornibrook2018creaky}). 
It appears to be more prevalent in females than in males (\cite{yuasa2010creaky},  \cite{wolk2012habitual}, \cite{abdelli-berus2014prevalence}). Several studies have reported high prevalence of creak in different groups of Finnish speakers. For instance, Pirilä et al.(\cite{pirila2017relationship}) observed that up to 54\% of teachers’ (N = 24) speech consisted of creaky voice. According to Ketolainen et al. (\cite{ketolainen2017speaking}), 60\% of 16-17 year old males and all except for one of the studied females (N= 40 in both groups) used a notable amount of creaky voice. According to \cite{laukkanen2021relations},  73.2 \% of female university students (N = 104) used slight to moderate amount of creaky voice. A recent study (\cite{uusitalo2022has}) reported that the prevalence of creaky voice use among Finnish female university students has increased significantly from the 1990's to 2010's, while among male students the change was not significant. However, in both groups, a tendency for increase in creaky voice use was observed. More specifically, it was found that the number of speakers who were perceptually rated as having ‘a lot of creak’ or ‘quite a lot of creak’ in their speech had increased from 5.9\% to 20.4\% in males and from 7\% to 31.8\% in females. In all the previous studies (\cite{laukkanen2021relations}, \cite{yuasa2010creaky}, \cite{hornibrook2018creaky}, \cite{wolk2012habitual} -- \cite{uusitalo2022has}), the amount of creak has been quantified either solely or primarily based on auditory perception.

Creaky voice in its different forms has several similarities with dysphonia: irregularity, super low pitch and characteristics of tense voice with a long closed phase and low pulse amplitude \cite{keating2015acoustic}. Dysphonia refers to vocal impairment as recognized by a clinician. According to the World Health Organization, an impairment is “any loss or abnormality of psychological, physiological or anatomical structure or function.” Characteristics of dysphonia include hoarseness, breathiness, voice fatigue, decreased vocal volume or other disturbances that limit the person’s performance in work-related and social functions \cite{sataloff2006voice}. Therefore it is challenging to distinguish ‘habitual creaky voice’ (i.e., social creak or voluntary dysphonia) from real dysphonia, either perceptually or with traditional acoustic analysis methods. Perceptual voice quality assessment in general is challenging  due to many reasons (\cite{kreiman2000sources}). A key reason is that it is difficult to extract and scale single  perceptual dimensions from the complex stimuli of natural voice and speech. Voice quality rating is particularly challenging for normophonic speakers, whose voice quality differences are more subtle than the differences between dysphonic and normophonic voices. For instance, according to \cite{faham2021acoustic}, there was low inter-rater reliability and low to moderate intra-rater reliability among voice specialists in rating students' voices for screening purposes. Ratings are affected by differences in individual scaling (e.g., \cite{kreiman2000sources}, \cite{bele2005reliability}). For instance, rating of the amount of creak in a speech sample may be affected by the number and duration of the creaky parts as well as the perceptual prominence of those parts (e.g., how loud they are or in which parts of the sentence or the whole sample they appear). Furthermore, cultural differences in voice quality profiles affect the rating. For instance, Italian voice experts gave lower scores for roughness than their French colleagues (\cite{ghio2015is}), which was explained by the higher prevalence of rough voice use in Italian speakers. Listeners in a study by Davidson (\cite{davidson2019perceptual}) were not able to distinguish between prototypical (fairly periodic) vocal fry and multiply-pulsed (more irregular) creaky voice. This seems to suggest that differentiation between dysphonic roughness and social creakiness may also be very difficult.  The uncertainty in the perceptual rating of normophonic speakers' voice characteristics calls for the use of more objective methods. However, particularly non-modal voices pose a challenge for acoustic analysis. For example, previous studies have shown that it is not possible to distinguish roughness, breathiness and vocal fry using  harmonic-to-noise ratio (HNR) (\cite{eskanazi1990acoustic}, \cite{dekrom1995some}). Furthermore, Eskenazy et al. (\cite{eskanazi1990acoustic}) observed that none of the studied acoustic measures (spectral flatness of the residue signal, pitch amplitude, HNR, pitch perturbation quotient, amplitude perturbation quotient, and percent jitter) was capable of ranking the normal voices. In a study by Laukkanen and Rantala (\cite{laukkanen2021relations}) there was, however, a low but statistically significant negative correlation between the perceived amount of creaky voice and HNR and smoothed cepstral peak prominence (CPPS) in healthy speakers. CPPS was also able to distinguish between samples evaluated as having low and high amount of creaky voice, although in the samples studied, the CPPS values were clearly higher (and thus better) than those related to dysphonic voices in earlier studies (e.g. \cite{sauder2017predicting}). Although cepstral measures are regarded as more reliable in the analysis of non-modal voices than the more traditional measures of jitter, shimmer and HNR (\cite{heman-ackah2003cepstral}), cepstral measures also have challenges: higher (and thus better) values may be obtained for hyperfunctional and louder voices and for a rough voice with a strong subharmonic component (\cite{awan2012effects},  \cite{awanawan2020two-stage}, \cite{brockmann-bauser2019effects}).

In addition to studying creaky voice perceptually and with traditional acoustical analysis based on signal parameters as described above, $automatic$ $detection$ of creaky voice has been also pursued in several studies over the past couple of decades. In this research line, the oldest category of previous studies is represented by investigations in which the topic has been addressed using a straightforward signal processing point of view by proposing acoustical measures of, for example, aperiodicity and using straightforward empirical rules based on the developed measures to detect creak (e.g., \cite{ishi2008amethod}, \cite{vishnubhotla2006automatic}). A more recent approach is represented by studies in which acoustical measures (often called hand-crafted features) have been used together with machine learning (ML) classifiers to train automatic pipeline systems to detect creaky voice. In this category, acoustical measures such as epoch parameters (\cite{narendra2015automatic}), the presence of secondary peaks and impulse-like excitation peaks in linear prediction residuals (\cite{drugman2012resonator-based}, \cite{kane2013improved}), as well as mel-frequency cepstral coefficients (MFCCs) and glottal features (\cite{borsky2017modal}) have been used as voice features. As ML classifiers, several known models such as support vector machines (SVMs) (\cite{surana2006acoustic}), binary decision tree (BDTs) and artificial neural networks (ANNs) (\cite{drugman2014datadriven}) have been studied. In addition, in some of the most recent studies, a deep learning (DL) based end-to-end approach has been used in the detection task. In end-to-end systems, the classical pipeline system consisting of two separate components (feature extraction and ML-based classifier) is replaced with one neural network which is optimized as a whole in the training phase. Examples of studies using the end-to-end approach include Chernyak et al. (\cite{chernyak2022deepfry}), which used a CNN-based network based on raw voice waveform inputs, and Chanclu et al. (\cite{chanclu2021automatic}), which used an autoencoder network.

In this study, we investigate the automatic classification of creaky voice using ML. Our focus is on social creak by studying voices of young female speakers of Finnish who have been classified as healthy, based on expert listeners' rating and self-perception, but who represent a talker category (\cite{uusitalo2022has}) which has shown a tendency to use creaky voice in their everyday speech communication. The goals of the study are as follows. First, by using recordings of continuous speech, we aim to understand whether automatic ML-based methods can be used to distinguish speakers who use low $vs$. high amount of creaky voice. Second, by comparing several potential existing ML classifiers, we aim to find the one that shows the best accuracy in the classification problem. The best classifier could then be used as a reference system in future studies where new detection systems are developed. Compared to the existing ML-based studies on creaky voice, the novelty of the current work can be summarized as follows. First, unlike most previous studies cited above, we study the classification problem using continuous (running) speech and by involving a larger number of speakers (N=90, as will be described in Section 2). Second, in all previous studies on automatic detection of creaky voice, the problem has been addressed as an automatic classification task between two classes (modal $vs$. creaky) by assuming that the modal class is free of creak. In the current study, we propose a more challenging task by assuming that $both$ classes may include creaky voice but the amount of creak is different (low $vs$. high). 
We argue that using continuous speech of multiple speakers as evaluation data and postulating the detection as a two-class problem described above are important generalizations in developing automatic ML tools that could potentially be used as alternatives to the time-consuming and often unreliable and culture-dependent subjective perceptual evaluations 
of social creak. Such tools would have applicability to sociolinguistic and vocological research and practice, including for voice screening and  training purposes.

\section{Data}
The speech data studied in this article are from the sound archive of Speech and Voice Research Laboratory at Tampere University (Finland). This data 
consist of continuous speech samples recorded in a text reading task from 118 female university students (mean age 23.7 years, SD 3.3 years, range 19-35 years). All participants read the same prose text extract of 158 words (in Finnish). Digital recordings were made in a recording studio using a sampling rate of 44.1 kHz and an amplitude resolution of 16 bits. 
In the samples recorded in the 1990’s, the microphone used was Bruel \& Kjaer 4165 and in the 2010’s it was Bruel \& Kjaer 4188. More details of the data can be found in \cite{uusitalo2022has}. 
The mouth-to-microphone distance was 40 cm. Normophonia of the speakers was checked by two speech therapists in a preliminary perceptual analysis of the text reading samples. Two voice specialists rated the amount of creak in the recorded speech samples. A 9-point Likert scale from 0 to 4 with a step of 0.5 was used to assess the amount of creak (0 = no creak at all, 4 = a lot of creak). The inter- and intra-rater reliabilities were satisfactory (Spearman’s rho 0.767, p $<$0.001, and rho 0.540, p $<$0.001, respectively). Therefore, means were calculated of the two raters' evaluations of creak. 
In order to get training and testing data for the proposed two-class classification problem, 
the samples rated by the two specialists were divided into the following two classes: (1) samples with low amount of social creak (samples with their mean Likert scale value $<$2) and (2) samples with high amount of social creak (samples with their mean Likert scale value $>$2). The number of samples in the two classes was balanced for the ML experiments, which finally yielded 45 samples of continuous speech for both classes (i.e., from the recorded set of 118 samples, 90 samples were included in this study).

\section{Machine learning-based classification}
Several ML-based systems were built in order to automatically classify speech into the two classes of social creak described in Section 2. All the systems used a three-stage architecture shown in Fig. 1 consisting of a pre-processing stage, feature extraction stage and classifier stage. 
In the system training, a classifier is trained using features extracted from the input speech sample and the binary label (low $vs$. high amount of social creak) of the sample. In the system testing, the same acoustic feature 
is computed from the input speech sample and fed to the trained classifier, which finally outputs the predicted class. In the current study, we compared three different features (to be described in Section 3.2) and seven different classifiers (to be described in Section 3.3), resulting in a total of 21 different automatic ML-based systems.

\begin{figure}[h]
\centering
\includegraphics[width=0.98\columnwidth,height=8.8cm]{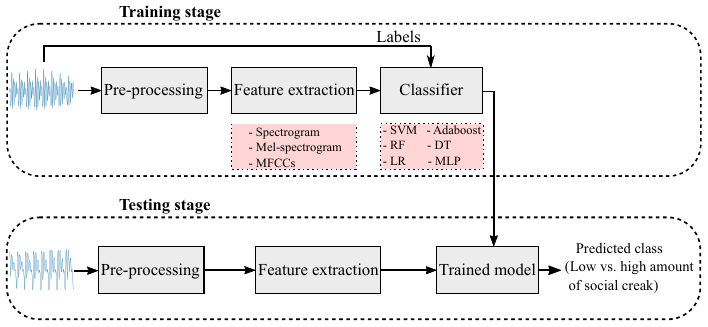}
\vspace{-0.5cm}
\caption{\label{fig:framework} A schematic block diagram of the training and testing stages of the ML system to automatically classify the amount of social creak in speech into two classes (low $vs$. high amount of social creak).
}
\label{fig:blockdiagram}
\end{figure}

\subsection{Pre-processing}
In the current study, a speech sample corresponds to a recorded signal of continuous speech. Therefore, the recorded raw speech sample includes sections of silence between sentences and words, and it needs to be pre-processed before it can be used in the system training and testing. In the first pre-processing step, the input speech sample is normalized by dividing the time domain signal waveform by its highest amplitude value. Subsequently, segments of silence are removed from the processed normalized speech sample using the sound exchange (SoX) tool \cite{barras2012sox}.
\subsection{Feature Extraction}
Next, the pre-processed input speech sample is converted into an acoustic feature vector. Three widely used spectral features (spectrogram, mel-spectrogram, and MFCCs) are extracted and compared in this study. These features have been shown to provide promising results in various tasks (e.g., automatic speech recognition \cite{povey2011kaldi}, speaker recognition \cite{snyder2018x}, and the classification of voice qualities \cite{borsky2017modal,kadiri2020analysis}). All these features are computed in time frames by using a frame length of 100 ms and a frame shift of 5 ms. These settings were selected based on our previous study \cite{tirronen2022effect}, which indicated that classification accuracy improves by using a frame length that is longer (e.g., 100 ms) than the widely-used frame length of a few tens milliseconds (e.g., 20 ms).  
For the computation of the spectrogram feature, speech is windowed frame-wise using a Hamming window. 
The amplitude spectrum is then estimated using a 1024-point fast Fourier transform (FFT). Subsequently, the 
logarithm of the amplitude spectrum is computed to get a 513-dimensional feature vector for each time frame. To derive the mel-spectrogram feature, the amplitude spectrum of the input speech sample is passed through an 128-channel mel-filterbank. The resulting mel-spectrogram is then transformed into the logarithmic decibel scale. This process yields a 128-dimensional feature vector for each time frame. The computation of the MFCCs involves employing the discrete cosine transform (DCT) on the mel-scale spectrum. From the resultant mel-cepstrum, first 13-cepstral coefficients (excluding the $0^{th}$ coefficient) are considered. In addition to the static coefficients, their first and second derivatives are calculated, which results in a 39-dimensional feature vector for each time frame. 

For all three feature extraction methods described above, the frame-wise computed features are merged into sample-wise features using statistical functionals as in \cite{tirronen2022effect,kadiri2019mel}. This is conducted using altogether eight different functionals (mean, standard deviation, median, skewness, kurtosis, minimum, maximum, and range). After this stage, the final feature dimension per speech sample is \( 513 \times 8 = 4104 \), \( 128 \times 8 = 1024 \), and \( 39 \times 8 = 312 \) for the spectrogram, mel-spectrogram and MFCCs, respectively.

\subsection{Classifiers}
Altogether seven different ML-based classifiers are compared: 
Support Vector Machine with linear kernel (SVM-linear), SVM with radial basis function kernel (SVM-RBF), Random Forest (RF), Multilayer Perceptron (MLP), Logistic Regression (LR), Decision Tree (DT), and Adaboost. All the ML classifiers were implemented using the Scikit-learn library \cite{scikit-learn}. 

The hyperparameters of the ML classifiers are optimized using a grid search strategy where a set of possible parameter combinations are first formed for each of the classifiers and the optimal combination is then searched. The parameters of the grid search are presented in Table \ref{tab:hyperparam}.  

\subsection{Evaluation}
To evaluate the performance of the ML models, the leave-one-speaker-out (LOSO) cross-validation scheme is employed. During each iteration, the data of one speaker is designated as the testing dataset, while the remaining speakers' data are used to train the classifiers. To standardize the training and testing datasets, z-score normalization is applied, utilizing the mean and standard deviation values derived from the training data. The evaluation metric (accuracy) is averaged across all iterations to represent the model's overall performance.

\begin{table}[ht]
\centering
\caption{The hyperparameters of the seven ML classifiers.}
\label{tab:hyperparam}
\begin{tabular}{|l |l|} \hline 
{\bf Classifier} & {\bf Hyperparameters} \\ \hline \hline 
SVM (linear) & C=1.0 \\ \hline 
SVM (RBF) & C=1.0, gamma=0.1 \\ \hline 
Random Forest (RF) & n\_estimators=100, max\_depth=None, random\_state=0 \\ \hline 
Multilayer Perceptron (MLP) & solver=adam, alpha=0.01, hidden\_layer\_sizes=(100,) \\ \hline 
Logistic Regression (LR) & solver=lbfgs, C=1.0 \\ \hline 
Decision Tree (DT) & max\_depth=5 \\ \hline 
AdaBoost & n\_estimators=100, learning\_rate=1.0 \\ \hline

\end{tabular}

\end{table}

\section{Results}
This section reports results of the conducted automatic classification experiments. extraction. 
In Table 2, the results are reported in terms of accuracy (mean and standard deviation computed over the LOSO iterations) for each feature and classifier. 
The table also shows the results averaged over all features for each classifier and averaged over all classifiers for each feature. For easier interpretation of the results, we have illustrated the mean classification accuracy (in \%) for each feature and classifier in Figure 2.

\begin{figure}[h]
\centering
\includegraphics[width=\columnwidth, trim=0cm 0cm 0cm 0cm,clip]
{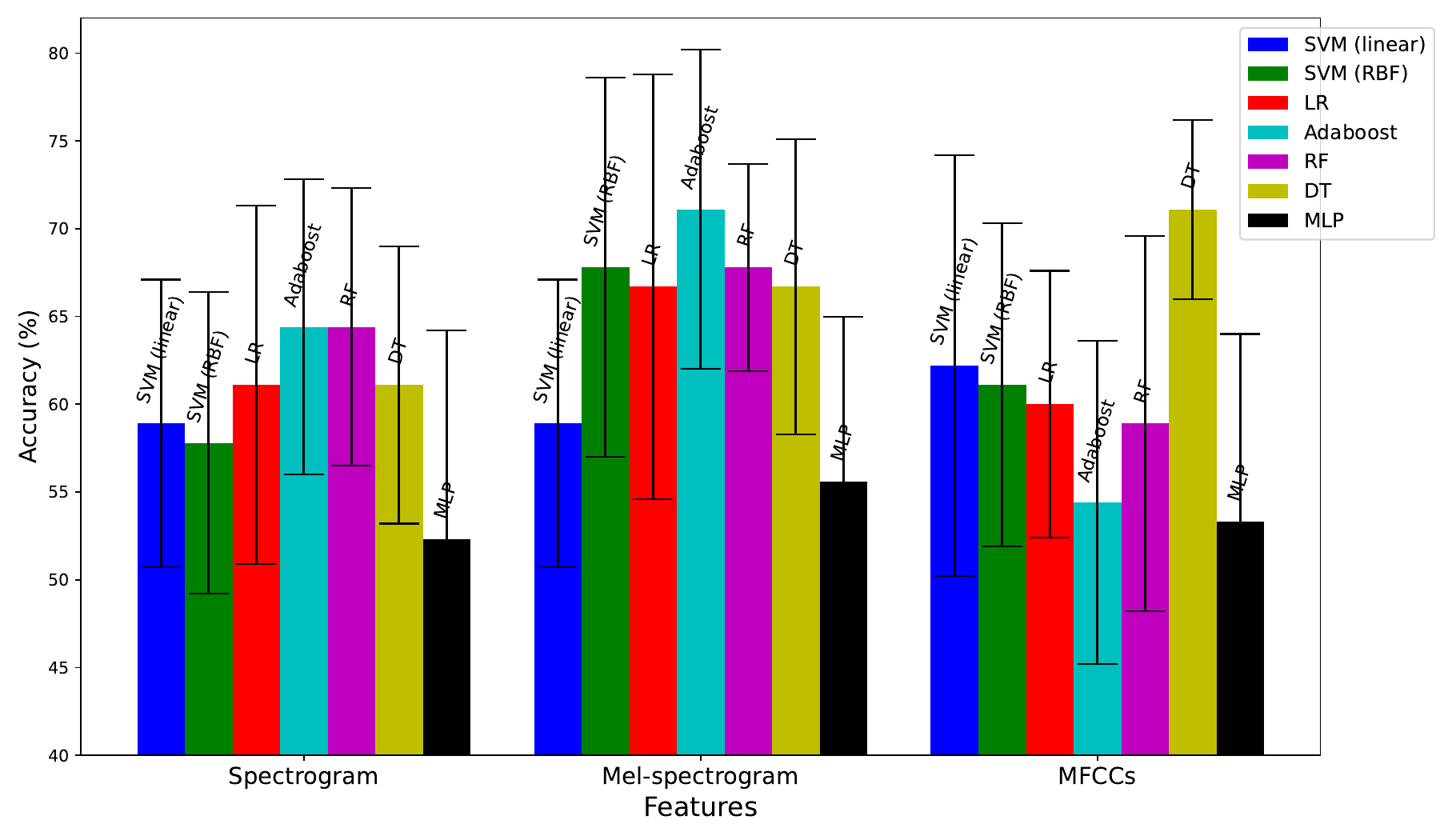}
\vspace{-1cm}
\caption{Mean accuracy (\%) along with standard deviation computed over the leave-one-speaker-out (LOSO) iterations for the three features (spectrogram, mel-spectrogram and MFCCs) and for the seven classifiers (Support Vector Machine (SVM) with a linear kernel and Radial Basis Function (RBF) kernel, Logistic Regression (LR), AdaBoost, Random Forest (RF), Decision Tree (DT), and Multilayer Perceptron (MLP)).}
\label{fig:longer-time}

\end{figure}

The values reported in Table 2 show that the best (mean) classification accuracy of 71.1\% was obtained by two systems: the Adaboost classifier, which used the mel-spectrogram feature, and the DT classifier, which used the MFCC feature. By comparing the three different spectral features for all classifiers using the averaged accuracy values reported on the lowest row of Table 2, it can be observed that the mel-spectrogram gives better performance compared to the spectrogram and MFCCs. The accuracy given by the mel-spectrogram feature was $>$65\% for all the compared classifiers except for SVM-lin and MLP. The classifier-wise averaged accuracy obtained using the spectrogram feature was similar to that of the MFCC feature. However, the accuracy obtained using the MFCC feature varied much more between the classifiers. In particular, using the MFCC feature for the DT classifier resulted in a clear improvement in classification accuracy compared with the other six classifiers as shown by the rightmost bars of Figure 2. 
Between the seven compared classifiers, DT stood out as the best classifier both in terms of feature-wise averaged accuracy (66.3\% in Table 2) and in terms of its accuracy when combined with the best feature (71.1\% in Table 2). As shown in Figure 2, the MLP classifier gave the lowest accuracy for all three feature representations.

\begin{table}[ht]
\centering
\caption{Accuracy (in \%) of the classification systems (mean $ \pm $ standard deviation computed over the LOSO iterations). Accuracy is shown on columns feature-wise and on rows classifier-wise. The best feature and classifier combination is printed in bold.}
\label{res:longer-time}
\vspace{0.1cm}

\begin{tabular}{|l|l|l|l||l|}
\hline
             & Spectrogram & Mel-spectrogram & MFCCs & Average over features\\ \hline \hline
SVM (linear) &58.9$\pm$8.2  &58.9$\pm$8.2      & 62.2$\pm$12.0   & 60.0  \\ \hline
SVM (RBF)    &57.8$\pm$8.6  &67.8$\pm$10.8      & 61.1$\pm$9.2   & 62.2  \\ \hline
LR           &61.1$\pm$10.2 &66.7$\pm$12.1      & 60.0$\pm$7.6   & 62.6  \\ \hline
Adaboost     &64.4$\pm$8.4  &{\bf71.1}$\pm$9.1      & 54.4$\pm$9.2   & 63.3  \\ \hline
RF           &64.4$\pm$7.9  &67.8$\pm$5.9      & 58.9$\pm$10.7   & 63.7  \\ \hline
DT           &61.1$\pm$7.9  &66.7$\pm$8.4      & {\bf71.1}$\pm$5.11   & 66.3 \\ \hline
MLP          &52.3$\pm$11.9 & 55.6$\pm$9.4     & 53.3$\pm$10.7   & 53.7  \\ \hline \hline
Average over classifiers  & 60.0  & 64.9  & 60.1   &  -- \\ \hline

\end{tabular}
\end{table}

\section{Discussion and Conclusions}
Several studies (e.g., \cite{yuasa2010creaky}, \cite{hornibrook2018creaky}, \cite{uusitalo2022has}) have reported increased prevalence of social creak especially among female speakers. Previous studies on social creak are based on conventional auditory-perceptual evaluation where the amount of creak in speech is assessed by voice specialists. Since such subjective evaluations are time-consuming, costly and may be subject to personal biases of raters, the use of automatic, ML-based assessment is an attractive new approach to study social creak. In the current study, ML-based approaches were developed to automatically classify the amount of social creak from continuous speech signals. We investigated a binary classification problem to distinguish speech signals into two categories of social creak (speech of low amount of social creak $vs$. speech of high amount of social creak). We built several ML-based systems based on supervised learning where an ML-based classifier was first trained using speech signals and their binary labels (low amount of social creak $vs$. high amount of social creak), which were obtained by averaging the amount of perceived creak rated on a 9-point Likert scale. 
All the systems were based on a three-stage architecture consisting of a pre-processing stage, a feature extraction stage and a classifier stage. We trained and tested altogether 21 systems by using three popular spectral feature representations and seven known ML models as the classifier.

The study showed that there were large differences in accuracy between the compared systems. The worst system (based on the MLP classifier and the spectrogram feature) gave an accuracy of 52.3\%, which is just barely above the chance level 50\% while the two best machines (based either on the Adaboost classifier and the mel-spectrogram feature or the DT classifier and the MFCC feature) were able to classify speech signals into the two categories with a clearly better accuracy of 71.1\%. As a general observation on the compared three features, the study found that the mel-spectrogram feature gave the highest average accuracy when considering all seven classifiers. The better performance of the mel-spectrogram feature is most likely due to the following issues that distinguish it from the other two feature representations used in the study. First, compared to the spectrogram feature, the mel-spectrogram feature models speech perception better by taking  advantage of mel-weighting of speech frequencies. Second, compared to the MFCC feature, which also uses mel-weighting but additionally compresses the spectrum extensively into a short vector, the mel-spectrogram feature is capable of carrying more spectral information of speech signals. Given that acoustical cues carrying information about social creak are often subtle making distinguishing the low and high amounts of social creak difficult even for human raters, we argue that mel-spectrogram's capability to combine perceptual weighting while also carrying adequate spectral redundancy makes it an effective feature for the studied ML-based task. Among the compared 21 systems, there was, however, one system which used the compressed MFCC feature representation but which still succeeded to show high accuracy. This system (shown by the second right-most bar in Figure 2) used the MFCC feature together with the DT classifier. The clearly higher accuracy of this system compared to the other six MFCC-based systems was most likely due to the DT classifier's ability to effectively leverage the discriminative power of the MFCC features. This suggests that while feature choice is critical, the selection of a classifier can enhance overall system performance.

In conclusion, the current study indicated that a machine algorithm is capable of automatically distinguishing speech of low amount of social creak from speech of high amount of social creak with a reasonable accuracy of about  70\%. We believe that this machine performance reported in the current study could serve as the baseline in future ML-based studies 
on social creak. In future studies, the current preliminary investigation could be extended in several ways by studying, for example, whether machine algorithms can also distinguish social creak from spontaneous speech and by comparing  performance between ML-based and deep learning -based classification models.
\section*{Acknowledgements}
This study was funded by the Academy of Finland (project no. 330139) and National Science Foundation (Grant: 2311676).


\begin{thebibliography}{10}
\expandafter\ifx\csname url\endcsname\relax
  \def\url#1{\texttt{#1}}\fi
\expandafter\ifx\csname urlprefix\endcsname\relax\def\urlprefix{URL }\fi
\expandafter\ifx\csname href\endcsname\relax
  \def\href#1#2{#2} \def\path#1{#1}\fi

\bibitem{keating2015acoustic}
P.~Keating, M.~Garellek, J.~Kreiman, Acoustic properties of different kinds of creaky voice, in: Proc. International Congress of Phonetic Sciences, 2015, pp. 2--7, paper no. 821.

\bibitem{hirano1981clincal}
M.~Hirano, Clinical Examination of Voice. Disorders of Human Communication, Springer Verlag, 1981.

\bibitem{dejonckere2001basic}
P.~Dejonckere, P.~Bradley, P.~Clemente, G.~Cornut, L.~Crevier-Buchman, G.~Friedrich, P.~{Van De Heyning}, M.~Remacle, V.~Woicard, A basic protocol for functional assessment of voice pathology, especially for investigating the efficacy of (phonosurgical) treatments and evaluating new assessment techniques. \uppercase{G}uideline elaborated by the \uppercase{C}ommittee on \uppercase{P}honiatrics of the \uppercase{E}uropean \uppercase{L}aryngological \uppercase{S}ociety (\uppercase{ELS}), Eur Arch Otorhinolaryngol 258 (2001) 77--82.

\bibitem{laukkanen2021relations}
A.-M. Laukkanen, L.~Rantala, Relations between creaky voice and vocal symptoms of fatigue, Folia Phoniatrica et Logopaedica 73 (2021) 146--154.

\bibitem{kreiman1982perception}
J.~Kreiman, Perception of sentence and paragraph boundaries in natural conversation, Journal of Phonetics 10~(2) (1982) 163--175.

\bibitem{henton1988creak}
C.~Henton, A.~Bladon, Creak as a socio-phonetic marker, in: L. Hyman and C. Li (Eds.), Language, Speech, and Mind (pp. 3–29). London: Routledge (1988).

\bibitem{redi2001variation}
L.~Redi, S.~Shattuck-Hufnagel, Variation in the realization of glottalization in normal speakers, Journal of Phonetics 29~(4) (2001) 407--429.

\bibitem{ogden2001turn}
R.~Ogden, Turn-holding, turn-yielding, and laryngeal activity in \uppercase{F}innish talk-in-interaction, Journal of the International Phonetic Association 31 (2001) 139--152.

\bibitem{abercrombie1971elements}
D.~Abercrombie, \uppercase{E}lements in \uppercase{G}eneral \uppercase{P}honetics, Edinburgh University Press, 1971.

\bibitem{laver1994principles}
J.~Laver, Principles of Phonetics, Cambridge University Press, 1994.

\bibitem{gordon2001phonation}
M.~Gordon, P.~Ladefoged, Phonation types: A cross-linguistic overview, Journal of Phonetics 29 (2001) 383--406.

\bibitem{cantor2017vocal}
L.~Cantor-Cutiva, P.~Bottalico, C.~Ishi, E.~Hunter, Vocal fry and vowel height in simulated room acoustics, Folia Phoniatrica et Logopaedica 69~(3) (2017) 118--124.

\bibitem{gobl2003role}
C.~Gobl, A.~NiChasaide, The role of voice quality in communicating emotion, mood and attitude, Speech Communication 40~(1-2) (2003) 189--212.

\bibitem{cullen2013creaky}
A.~Cullen, J.~Kane, T.~Drugman, N.~Harte, Creaky voice and the classification of affect, in: Proc. Workshop on Affective Social Speech Signals, 2013, pp. 1--5.

\bibitem{oliveira2016comparison}
G.~Oliveira, A.~Davidson, R.~Holczer, S.~Kaplan, A.~Paretzky, A comparison of the use of glottal fry in the spontaneous speech of young and middle-aged \uppercase{A}merican women, Journal of Voice 30~(6) (2016) 684--687.

\bibitem{yuasa2010creaky}
I.~Yuasa, Creaky voice: A new feminine voice quality for young urban-oriented upwardly mobile \uppercase{A}merican women?, American Speech 85~(3) (2010) 315–237.

\bibitem{mendoza2011semiotic}
N.~Mendoza-Denton, The semiotic hitchhiker’s guide to creaky voice: \uppercase{C}irculation and gendered hardcore in a \uppercase{C}hicana/o gang persona, Journal of Linguistic Anthropology 21 (2011) 261–280.

\bibitem{aare2014creaky}
K.~Aare, P.~Lippus, J.~Simko, Creaky voice in spontaneous spoken \uppercase{E}stonian, in: Proceedings of the XXVIIIth Finnish Phonetics Symposium (Fonetiikan päivät) in Turku 2013, K. Jähi and L. Taimi (Eds.). University of Turku (2014).

\bibitem{blomgren1998acoustic}
M.~Blomgren, Y.~Chen, M.~Ng, H.~Gilbert, Acoustic, aerodynamic, physiologic, and perceptual properties of modal and vocal fry registers, Journal of the Acoustical Society of America 103~(5) (1998) 2649--2658.

\bibitem{hornibrook2018creaky}
J.~Hornibrook, T.~Ormond, M.~Maclagan, Creaky voice or extreme vocal fry in young women, The New Zealand Medical Journal 131 (2018) 36--40.

\bibitem{wolk2012habitual}
L.~Wolk, N.~Abdelli-Beruh, D.~Slavin, Habitual use of vocal fry in young adult female speakers, Journal of Voice 26~(3) (2012) e111--e116.

\bibitem{abdelli-berus2014prevalence}
N.~Abdelli-Beruh, L.~Wolk, D.~Slavin, Prevalence of vocal fry in young adult male \uppercase{A}merican \uppercase{E}nglish speakers, Journal of Voice 28~(2) (2014) 185--190.

\bibitem{pirila2017relationship}
S.~Pirilä, P.~Pirilä, T.~Ansamaa, A.~Yliherva, S.~Sonning, L.~Rantala, Relationship between activity noise, voice parameters, and voice symptoms among female teachers, Folia Phoniatrica et Logopaedica 69~(3) (2017) 94--102.

\bibitem{ketolainen2017speaking}
I.~Ketolainen, M.~Laakso, S.~Simberg, Speaking pitch in 16–17-year-old \uppercase{F}innish teenagers (in \uppercase{F}innish; 16-17-vuotiaiden suomalaisnuorten puheäänen korkeus), Puhe ja kieli 37 (2017) 259--277.

\bibitem{uusitalo2022has}
T.~Uusitalo, L.~Nyberg, A.-M. Laukkanen, T.~Waaramaa, L.~Rantala, Has the prevalence of creaky voice increased among {F}innish university students from the 1990's to the 2010's?, Journal of VoiceIn press.

\bibitem{sataloff2006voice}
R.~T. Sataloff, Voice impairment, disability, handicap and medical/legal evaluation, in: Diagnosis and Treatment of Voice Disorders, Plural Publishing, 2006, pp. 319--327.

\bibitem{kreiman2000sources}
J.~Kreiman, B.~Gerratt, Sources of listener disagreement in voice quality assessment, Journal of the Acoustical Society of America 108~(4) (2000) 1867--1876.

\bibitem{faham2021acoustic}
M.~Faham, A.-M. Laukkanen, T.~Ikävalko, L.~Rantala, A.~Geneid, S.~Holmqvist-Jämsen, K.~Ruusuvirta, S.~Pirilä, Acoustic voice quality index as a potential tool for voice screening, Journal of Voice 35~(2) (2021) 226--232.

\bibitem{bele2005reliability}
I.~Bele, Reliability in perceptual analysis of voice quality, Journal of Voice 19~(4) (2005) 555--573.

\bibitem{ghio2015is}
A.~Ghio, G.~Cantarella, F.~Weisz, D.~Robert, V.~Woisard, F.~Fussi, A.~Giovanni, G.~Baracca, Is the perception of dysphonia severity language-dependent? \uppercase{A} comparison of \uppercase{F}rench and \uppercase{I}talian voice assessments, Logopedics Phoniatrics and Vocology 40~(1) (2015) 36--43.

\bibitem{davidson2019perceptual}
L.~Davidson, Perceptual coherence of creaky voice qualities, in: Proc. International Congress of Phonetic Sciences, 2019, pp. 147--151.

\bibitem{eskanazi1990acoustic}
L.~Eskanazi, D.~Childers, D.~Hicks, Acoustic correlates of vocal quality, Journal of Speech and Hearing Research 33 (1990) 298--306.

\bibitem{dekrom1995some}
G.~{de Krom}, Some spectral correlates of pathological breathy and rough voice quality for different types of vowel fragments, Journal of Speech and Hearing Research 38 (1995) 794--811.

\bibitem{sauder2017predicting}
C.~Sauder, M.~Bretl, T.~Eadie, Predicting voice disorder status from smoothed measures of cepstral peak prominence using \uppercase{P}raat and \uppercase{A}nalysis of dysphonia in speech and voice (\uppercase{ADSV}), Journal of Voice 31~(5) (2017) 557--566.

\bibitem{heman-ackah2003cepstral}
J.~Heman-Ackah, D.~Michael, M.~Baroody, R.~Ostrowski, J.~Hillenbrand, R.~Heuer, M.~Horman, R.~Sataloff, Cepstral peak prominence: A more reliable measure of dysphonia, Annals of Otology, Rhinology and Laryngology 112~(4) (2003) 324--333.

\bibitem{awan2012effects}
S.~Awan, A.~Giovinco, J.~Owens, Effects of vocal intensity and vowel type on cepstral analysis of voice, Journal of Voice 26~(5) (2012) 670.e15--670.e20.

\bibitem{awanawan2020two-stage}
S.~Awan, J.~Awan, A two-stage cepstral analysisprocedure for the classification of rough voices, Journal of Voice 34~(1) (2020) 9--19.

\bibitem{brockmann-bauser2019effects}
M.~Brockmann-Bauser, J.~VanStan, M.~Sampaio, J.~Bohlender, R.~Hillman, D.~Mehta, Effects of vocal intensity and fundamental frequency on cepstral peak prominence in patients with voice disorders and vocally healthy controls, Journal of Voice 35~(3) (2019) 411--417.

\bibitem{ishi2008amethod}
C.~T. Ishi, K.-I. Sakakibara, H.~Ishiguro, N.~Hagita, A method for automatic detection of vocal fry, IEEE Transactions on Audio, Speech, and Language Processing 16~(1) (2008) 47--56.

\bibitem{vishnubhotla2006automatic}
S.~Vishnubhotla, C.~Espy-Wilson, Automatic detection of irregular phonation in continuous speech, in: Proc. Interspeech, 2006, pp. 949--952.

\bibitem{narendra2015automatic}
N.~P. Narendra, K.~S. Rao, Automatic detection of creaky voice using epoch parameters, in: Proc. Interspeech, 2015, pp. 2347--2351.

\bibitem{drugman2012resonator-based}
T.~Drugman, J.~Kane, C.~Gobl, Resonator-based creaky voice detection, in: Proc. Interspeech, 2012, pp. 1424--1427.

\bibitem{kane2013improved}
J.~Kane, T.~Drugman, C.~Gobl, Improved automatic detection of creak, Computer Speech and Language 27~(4) (2013) 1028--1047.

\bibitem{borsky2017modal}
M.~Borsky, D.~D. Mehta, J.~H. Van~Stan, J.~Gudnason, Modal and nonmodal voice quality classification using acoustic and electroglottographic features, IEEE/ACM transactions on audio, speech, and language processing 25~(12) (2017) 2281--2291.

\bibitem{surana2006acoustic}
K.~Surana, J.~Slifka, Acoustic cues for the classification of regular and irregular phonation, in: Proc. Interspeech, 2006, pp. 693--696.

\bibitem{drugman2014datadriven}
T.~Drugman, J.~Kane, C.~Gobl, Data-driven detection and analysis of the patterns of creaky voice, Computer Speech and Language 28~(5) (2014) 1233--1253.

\bibitem{chernyak2022deepfry}
B.~R. Chernyak, T.~{Ben Simon}, Y.~Segal, J.~Steffman, E.~Chodroff, J.~Cole, J.~Keshet, Deepfry: Identifying vocal fry using deep neural networks, in: Interspeech, 2022, pp. 3578--3582.

\bibitem{chanclu2021automatic}
A.~Chanclu, I.~B. Amor, C.~Gendrot, E.~Ferragne, J.-F. Bonastre, Automatic classification of phonation types in spontaneous speech: towards a new workflow for the characterization of speakers’ voice quality, in: Proc. Interspeech, 2021, pp. 1015--1018.

\bibitem{barras2012sox}
B.~Barras, Sox: Sound exchange, Tech. rep. (2012).

\bibitem{povey2011kaldi}
D.~Povey, A.~Ghoshal, G.~Boulianne, L.~Burget, O.~Glembek, N.~Goel, M.~Hannemann, P.~Motlicek, Y.~Qian, P.~Schwarz, et~al., The \uppercase{K}aldi speech recognition toolkit, in: Automatic speech recognition and understanding, IEEE Signal Processing Society, 2011.

\bibitem{snyder2018x}
D.~Snyder, D.~Garcia-Romero, G.~Sell, D.~Povey, S.~Khudanpur, X-vectors: Robust \uppercase{dnn} embeddings for speaker recognition, in: IEEE International Conference on Acoustics, Speech and Signal Processing (ICASSP), 2018, pp. 5329--5333.

\bibitem{kadiri2020analysis}
S.~R. Kadiri, P.~Alku, Analysis and classification of phonation types in speech and singing voice, Speech Communication 118 (2020) 33--47.

\bibitem{tirronen2022effect}
S.~Tirronen, S.~R. Kadiri, P.~Alku, The effect of the \uppercase{MFCC} frame length in automatic voice pathology detection, Journal of Voice 38~(5) (2024) 975--982.

\bibitem{kadiri2019mel}
S.~R. Kadiri, P.~Alku, Mel-frequency cepstral coefficients derived using the zero-time windowing spectrum for classification of phonation types in singing, The Journal of the Acoustical Society of America 146~(5) (2019) EL418--EL423.

\bibitem{scikit-learn}
F.~Pedregosa, G.~Varoquaux, A.~Gramfort, V.~Michel, B.~Thirion, O.~Grisel, M.~Blondel, P.~Prettenhofer, R.~Weiss, V.~Dubourg, J.~Vanderplas, A.~Passos, D.~Cournapeau, M.~Brucher, M.~Perrot, E.~Duchesnay, Scikit-learn: Machine learning in {P}ython, Journal of Machine Learning Research 12 (2011) 2825--2830.

\end{thebibliography}

\end{document}